\documentclass[apj,iop]{emulateapj}
\usepackage{color}
\usepackage{times}
\usepackage{hyperref}
\hypersetup{
    pdftitle={Volumetric Type Ia Supernova Rates}, 
    pdfauthor={Kyle Barbary},
    colorlinks=true,        
    linkcolor=blue,         
    citecolor=blue,         
    urlcolor=blue           
}
\newcommand\T{\rule{0pt}{2.6ex}}
\newcommand\B{\rule[-1.2ex]{0pt}{0pt}}

\slugcomment{Submitted to the Astrophysical Journal}
\shorttitle{Volumetric Type Ia Supernova Rates}
\shortauthors{Barbary et al.}
\begin{document}
\title{The {\it Hubble Space Telescope}\footnotemark[1] 
       Cluster Supernova Survey:\\
       VI. The Volumetric Type I\lowercase{a} 
       Supernova Rate}
\footnotetext[1]{Based in part on observations made with the NASA/ESA Hubble 
  Space Telescope, obtained from the data archive at the Space
  Telescope Institute. STScI is operated by the association of
  Universities for Research in Astronomy, Inc. under the NASA contract
  NAS 5-26555.  The observations are associated with program GO-10496.}
\author{
K.~Barbary\altaffilmark{2,3},
G.~Aldering\altaffilmark{3},
R.~Amanullah\altaffilmark{2,4},
M.~Brodwin\altaffilmark{5,6},
N.~Connolly\altaffilmark{7},
K.~S.~Dawson\altaffilmark{3,8},
M.~Doi\altaffilmark{9},
P.~Eisenhardt\altaffilmark{10},
L.~Faccioli\altaffilmark{3},
V.~Fadeyev\altaffilmark{11},
H.~K.~Fakhouri\altaffilmark{2,3},
A.~S.~Fruchter\altaffilmark{12},
D.~G.~Gilbank \altaffilmark{13},
M.~D.~Gladders\altaffilmark{14},
G.~Goldhaber\altaffilmark{2,3,27}, 
A.~Goobar\altaffilmark{4,15},
T.~Hattori\altaffilmark{16},
E.~Hsiao\altaffilmark{3},
X.~Huang\altaffilmark{2},
Y.~Ihara\altaffilmark{9,26},
N.~Kashikawa\altaffilmark{17},
B.~Koester\altaffilmark{14,18},
K.~Konishi\altaffilmark{19},
M.~Kowalski\altaffilmark{20},
C.~Lidman\altaffilmark{21},
L.~Lubin\altaffilmark{22},
J.~Meyers\altaffilmark{2,3},
T.~Morokuma\altaffilmark{9,17,26},
T.~Oda\altaffilmark{23},
N.~Panagia\altaffilmark{12},
S.~Perlmutter\altaffilmark{2,3},
M.~Postman\altaffilmark{12},
P.~Ripoche\altaffilmark{3},
P.~Rosati\altaffilmark{24},
D.~Rubin\altaffilmark{2,3},
D.~J.~Schlegel\altaffilmark{3},
A.~L.~Spadafora\altaffilmark{3},
S.~A.~Stanford\altaffilmark{22,25},
M.~Strovink\altaffilmark{2,3},
N.~Suzuki\altaffilmark{3},
N.~Takanashi\altaffilmark{17},
K.~Tokita\altaffilmark{9},
N.~Yasuda\altaffilmark{19} \\
(The Supernova Cosmology Project)
}
\email{kbarbary@lbl.gov}
\altaffiltext{2}{Department of Physics, University of California
Berkeley, Berkeley, CA 94720}
\altaffiltext{3}{E.O. Lawrence Berkeley National Lab, 1 Cyclotron Rd.,
Berkeley CA, 94720}
\altaffiltext{4}{The Oskar Klein Centre for Cosmo Particle Physics,
AlbaNova, SE-106 91 Stockholm, Sweden}
\altaffiltext{5}{Harvard-Smithsonian Center for Astrophysics, 60
Garden Street, Cambridge, MA 02138}
\altaffiltext{6}{W. M. Keck Postdoctoral Fellow at the
Harvard-Smithsonian Center for Astrophysics}
\altaffiltext{7}{Hamilton College Department of Physics, Clinton, NY 13323}
\altaffiltext{8}{Department of Physics and Astronomy, University of
Utah, Salt Lake City, UT 84112}
\altaffiltext{9}{Institute of Astronomy, Graduate School of Science,
University of Tokyo 2-21-1 Osawa, Mitaka, Tokyo 181-0015, Japan}
\altaffiltext{10}{Jet Propulsion Laboratory, California Institute of
Technology, Pasadena, CA, 91109}
\altaffiltext{11}{Santa Cruz Institute for Particle Physics,
University of California, Santa Cruz, CA 94064}
\altaffiltext{12}{Space Telescope Science Institute, 3700 San Martin
Drive, Baltimore, MD 21218, USA}
\altaffiltext{13}{Department of Physics and Astronomy, 
University Of Waterloo, Waterloo, Ontario, Canada N2L 3G1}
\altaffiltext{14}{Department of Astronomy and Astrophysics, University
of Chicago, Chicago, IL 60637}
\altaffiltext{15}{Department of Physics, Stockholm University,
Albanova University Center, SE-106 91, Stockholm, Sweden}
\altaffiltext{16}{Subaru Telescope, National Astronomical Observatory
of Japan, 650 North A'ohaku Place, Hilo, HI 96720}
\altaffiltext{17}{Institute of Industrial Science, The University of Tokyo,
 4-6-1 Komaba, Meguro-ku, Tokyo, 153-8505, Japan}
\altaffiltext{18}{Kavli Institute for Cosmological Physics, The University of
Chicago, Chicago IL 60637, USA}
\altaffiltext{19}{Institute for Cosmic Ray Research, University of
Tokyo, 5-1-5, Kashiwanoha, Kashiwa, Chiba, 277-8582, Japan}
\altaffiltext{20}{Physikalisches Institut, Universit\"at Bonn, Bonn, Germany}
\altaffiltext{21}{Australian Astronomical Observatory, PO Box 296, Epping, 
NSW 1710, Australia}
\altaffiltext{22}{University of California, Davis, CA 95618}
\altaffiltext{23}{Department of Astronomy, Kyoto University, Sakyo-ku,
Kyoto 606-8502, Japan}
\altaffiltext{24}{ESO, Karl-Schwarzschild-Strasse 2, D-85748 Garching, Germany}
\altaffiltext{25}{Institute of Geophysics and Planetary Physics,
Lawrence Livermore National Laboratory, Livermore, CA 94550}
\altaffiltext{26}{JSPS Fellow}
\altaffiltext{27}{Deceased}
\begin{abstract}

We present a measurement of the volumetric Type Ia supernova (SN~Ia)
rate out to $z \simeq 1.6$ from the \emph{Hubble Space Telescope}
Cluster Supernova Survey. In observations spanning 189 orbits with the
Advanced Camera for Surveys we discovered 29 SNe, of which
approximately 20 are SNe~Ia. Twelve of these SNe~Ia are located in the
foregrounds and backgrounds of the clusters targeted in the survey.
Using these new data, we derive the volumetric SN~Ia rate in four
broad redshift bins, finding results consistent with previous
measurements at $z \gtrsim 1$ and strengthening the case for a SN~Ia
rate that is $\gtrsim$$0.6 \times 10^{-4}
h_{70}^{3}$~yr$^{-1}$~Mpc$^{-3}$ at $z \sim 1$ and flattening out at
higher redshift. We provide SN candidates and efficiency calculations
in a form that makes it easy to rebin and combine these results with
other measurements for increased statistics. Finally, we compare the
assumptions about host-galaxy dust extinction used in different
high-redshift rate measurements, finding that different assumptions
may induce significant systematic differences between measurements.

\end{abstract}
\section{Introduction}

Type Ia supernovae (SNe~Ia) are of great importance both as
astrophysical objects and as cosmological distance indicators. An
accurate knowledge of the rate at which they occur (as a function of
redshift) is essential for understanding both of these roles.
Astrophysically, SNe~Ia play an important role in galaxy evolution.
They are a major source of
iron \citep[e.g.,][]{matteucci86a,tsujimoto95a,thielemann96a} and
inject energy into the interstellar
medium \citep[e.g.,][]{dekel86a,scannapieco06a}. The SN~Ia rate is
necessary to include these effects in galaxy evolution models,
particularly at high redshifts where much of the important galaxy
evolution occurs.  Cosmologically, SNe~Ia are the best-tested method
for measuring the scale factor of the universe as a function of
redshift, with hundreds of SNe now employed in the precision
measurement of cosmological
parameters \citep[e.g.,][]{hicken09a,amanullah10a,sullivan11a}. Despite
their widespread use as distance indicators, the process that leads to
a SN~Ia is still not well understood. SNe~Ia are widely accepted to be
the end result of a carbon-oxygen (CO) white dwarf (WD) nearing the
Chandrasekhar mass limit but how they near that limit is not
known \citep[see][for a review]{livio01a}. This leaves open the
question of whether high-redshift SNe are different from low-redshift
SNe in a way that affects the inferred distance.

Measurements of the change in the SN~Ia rate with redshift can be used
to distinguish between models of how SNe~Ia occur.  While there are a
variety of SN~Ia progenitor models, most fall into two classes: the
{\it single degenerate scenario} \citep[SD;][]{whelan73a} and 
the {\it double degenerate
scenario} \citep[DD;][]{iben84a,webbink84a}. In the single degenerate scenario,
the WD accretes mass from a red giant or main sequence star that
overflows its Roche lobe.  In the double degenerate scenario, the WD merges
with a second white dwarf after orbital decay due to the emission of
gravitational radiation. Crucially, the delay time between the initial
formation of the system and the SN explosion is governed by a
different physical mechanism in the different models. This allows us
to differentiate between models by measuring the distribution of the
delay times for a population of SNe.  The shape of this delay time
distribution (DTD) depends on the details of the binary star evolution
(particularly its evolution through one or more common envelope
phases) and the specific progenitor model. One method for measuring
the DTD is to correlate the cosmic star formation history (SFH) with
the the cosmic SN~Ia rate as a function of
redshift \citep{yungelson00a}: the rate as a function of cosmic time
is simply the cosmic SFH convolved with the DTD.

The volumetric SN~Ia rate has now been measured in many different SN
surveys designed to detect and measure SNe at
$z<1$ \citep[e.g.][]{pain02a,neill07a,dilday10a}. With the recently
revised rates from the IfA Deep survey \citep{rodney10a}, most of
these $z<1$ measurements have now come into agreement.  In contrast,
measurements at $z>1$ have been limited to SN searches in the
GOODS\footnote{Great Observatories Origins Deep
Survey \citep{giavalisco04a}}
fields \citep{dahlen04a,kuznetsova08a,dahlen08a} using
the \emph{Hubble Space Telescope} (\emph{HST}) and ultra-deep
single-epoch searches in the Subaru Deep Field (SDF) from the
ground \citep{poznanski07a,graur11a}. These studies have yielded
discrepant results for the
DTD. The first $z>1$ measurements by \citet{dahlen04a} \citep[and
later][with an expanded dataset]{dahlen08a} showed a rate that peaked
at $z \sim 1$ and decreased in the highest-redshift bin at
$z>1.4$. From these results the best-fit DTD is one tightly confined
to 3--4 Gyr with very few SNe Ia having short delay times \citep{strolger10a}.
The recent results of \citet{graur11a} from the SDF show a lower rate
at $z \sim 1$, and a higher rate in the highest-redshift bin compared
with \citet{dahlen08a}.  These results are consistent with a flat SN
rate at $1<z<2$. They find that the DTD is consistent with a power law
with the best-fit $\propto t^{-1.1}$, implying a significant fraction
of short delay time ($\lesssim 1$~Gyr) SNe.

Relative to the \emph{HST} measurements, the SDF measurements cover a
much larger volume and therefore have the advantage of better
statistics in the highest-redshift bin, but \emph{HST} measurements
hold advantages in systematics.  A rolling search with \emph{HST}
offers multiple observations of each SN and much higher resolution
than possible from the ground, useful for resolving SNe from the cores
of their hosts. These factors lead to a more robust identification of
SNe~Ia relative to the SDF searches where a single observation is used
for both detection and photometric typing. In addition,
the \citet{dahlen08a} analysis used spectroscopic typing in addition
to photometric typing, whereas \citet{graur11a} uses only photometric
typing. In general, the very different strategies employed
make \emph{HST} measurements a good cross-check for the SDF
measurements and vice versa. Increasing the statistics in \emph{HST}
rate measurements can make this cross-check better and improve DTD
constraints. At the same time, in comparing the measurements it is
important to carefully consider possible systematic differences,
particularly as statistical uncertainty decreases and systematics come
to dominate.

In this paper, we address these issues by (1) supplementing current
determinations of the \emph{HST}-based $z \gtrsim 1$ SN~Ia rate and (2)
comparing the effect on results of different dust distributions assumed in
previous analyses. We use observations from the \emph{HST} Cluster Supernova
Survey, a survey to discover and follow SNe~Ia in very distant clusters
\citep[][PI: Perlmutter, GO-10496]{dawson09a}. The survey encompassed 189 orbits
with the Advanced Camera for Surveys (ACS) over a period of 18 months. The SN
selection and SN typing for the all SNe in the survey was presented in
\citet[][hereafter B11]{barbary11a}, where we calculated the \emph{cluster}
SN~Ia rate from the survey. Here, we use a similar methodology to B11 but focus
on the SNe discovered in the cluster foregrounds and backgrounds. The remainder
of the paper is organized as follows: In \S\ref{data} we summarize the
\emph{HST} Cluster Supernova Survey and the SN discoveries.  In \S\ref{ratecalc}
we describe the Monte Carlo simulation used to calculate a rate based
on the SN discoveries. In \S\ref{results} we present results and
characterize systematic uncertainties. Finally, in \S\ref{discussion}
we compare our results to published measurements.  Throughout the paper we
use a cosmology with $H_0=70$~km~s$^{-1}$~Mpc$^{-1}$, $\Omega_M=0.3$,
$\Omega_\Lambda = 0.7$. Magnitudes are in the Vega system.

This paper is one of a series of ten papers \citep[][B11; This
work]{melbourne07a,barbary09a,dawson09a,morokuma10a,suzuki11a,ripoche11a,meyers11a,hsiao11a}
that report supernova results from the \emph{HST} Cluster Supernova
Survey. The survey strategy and SN discoveries are described
in \citet{dawson09a}, while spectroscopic follow-up observations for
SN candidates are presented in \citet{morokuma10a}. A separate series
of papers, ten to date, reports on cluster studies from the survey:
\citet{hilton07a,eisenhardt08a,jee09a,hilton09a,huang09a,rosati09a,
santos09a,strazzullo10a,brodwin10a,jee11a}.

\section{Survey and Supernova Discoveries} \label{data}

The details of the \emph{HST} Cluster SN Survey are described in
\citet{dawson09a}.  Briefly, the survey targeted 25 massive galaxy 
clusters in a rolling SN search between July 2005 and December 2006.
Clusters were selected from X-ray, optical and IR surveys and cover
the redshift range $0.9<z<1.46$.  During the survey, each cluster was
observed once every 20 to 26 days during its \emph{HST} visibility
window (typically four to seven months) using ACS. Each visit
consisted of four exposures in the F850LP filter (hereafter
$z_{850}$).  Most visits also included a fifth exposure in the F775W
filter (hereafter $i_{775}$).

\begin{deluxetable}{lccccc}
\tablewidth{\columnwidth}
\tablecaption{\label{table:sne}Non-Cluster Supernova Discoveries}
\tablehead{\colhead{ID} & \colhead{Nickname} & \colhead{$z$} & 
           \colhead{Cluster $z$} & \colhead{Type} & \colhead{Confidence}}
\startdata
\sidehead{\emph{SNe: Not in Clusters}}
SN SCP06L21\tablenotemark{a} & \nodata   & \nodata & 1.37 & CC      & plausible\\
SN SCP05N10\tablenotemark{a} & Tobias    & 0.203   & 1.03 & CC      & plausible\\
SN SCP06C7  & \nodata   & 0.61\phn & 0.97 & CC      & probable \\
SN SCP06Z5  & Adrian    & 0.623    & 1.39 & Ia      & secure\tablenotemark{b}   \\
SN SCP06B3\tablenotemark{c} & Isabella  & 0.743   & 1.12 & CC      & probable \\
SN SCP06F8  & Ayako     & 0.789    & 1.11 & CC      & probable \\
SN SCP05P9  & Lauren    & 0.821    & 1.1\phn  & Ia      & secure\tablenotemark{b}   \\
SN SCP06H3  & Elizabeth & 0.85\phn & 1.24 & Ia      & secure\tablenotemark{b}   \\
SN SCP06U7  & Ingvar    & 0.892    & 1.04 & CC      & probable \\
SN SCP05P1  & Gabe      & 0.926    & 1.1\phn & Ia      & probable \\
SN SCP06G3  & Brian     & 0.962    & 1.26 & Ia      & plausible\\
SN SCP06C0  & Noa       & 1.092    & 0.97 & Ia      & secure   \\
SN SCP06N33 & Naima     & 1.188    & 1.03 & Ia      & probable \\
SN SCP06F6  & \nodata   & 1.189    & 1.11 & non-Ia  & secure   \\
SN SCP06A4  & Aki       & 1.193    & 1.46 & Ia      & probable \\
SN SCP05D6  & Maggie    & 1.314    & 1.02 & Ia      & secure   \\
SN SCP06G4  & Shaya     & 1.35\phn & 1.26 & Ia      & secure\tablenotemark{b}   \\
SN SCP06X26 & Joe       & 1.44?\tablenotemark{d} & 1.10 & Ia  & plausible\\
\sidehead{\emph{SNe: Cluster Membership Uncertain}}
SN SCP06E12 & Ashley    & \nodata  & 1.03 & Ia      & plausible \\
SN SCP06N32 & \nodata   & \nodata  & 1.03 & CC      & plausible \\
\sidehead{\emph{Uncertain to be SN}}
SN SCP06M50\tablenotemark{c} & \nodata   & \nodata & 0.90 & \nodata & \nodata

\enddata
\tablenotetext{a}{Excluded from this analysis due to being inconsistent with an SN~Ia peaking $<10$~rest-frame days before the first observation. These SNe are excluded due to the difficulty of typing SNe found far on the decline.}
\tablenotetext{b}{Spectroscopically confirmed. Spectroscopy for SNe 
SCP05P9, SCP06H3 and SCP06G4 is reported in \citet{morokuma10a}}
\tablenotetext{c}{Excluded from this analysis due to being within $20''$ of cluster center. These regions are excluded to reduce complications of lensing.}
\tablenotetext{d}{Based on a marginal emission line at 9100~\AA{} \citep[see][]{morokuma10a}.}

\end{deluxetable}

The process of selecting SN~Ia candidates for the rates analysis is
described in detail in B11. We briefly summarize it here. Candidates
were detected in subtractions of $z_{850}$ images and examined by eye
to eliminate obviously false detections such as stellar diffraction
spikes. A total of 86 candidates were selected in this phase. Detailed
information on all 86 of these candidates is available from the survey
website\footnote{\url{http://supernova.lbl.gov/2009ClusterSurvey/}}. We
generated a full light curve in $z_{850}$ and $i_{775}$ for each of
these candidates and then imposed automated requirements on the light
curve. These included a requirement on the flux in $i_{775}$ and the
rapidity of the rise and fall of the light curve. After this step, 60
SN candidates remained. (The selections up to this point are accounted
for in our calculation of detection efficiency.) The remaining
candidates were then divided into image artifacts (14), AGN (17), and
supernovae (29) on the basis of the light curve shape and evidence
from image subtractions. For example, candidates on the cores of
bright galaxies and showing adjoining positive and negative regions in
image subtractions are likely to be the result of image
mis-alignment. With corroborating evidence from the light curves of
such candidates, they are confidently dismissed as image
artifacts. Similarly, candidates deemed to be AGN were located on the
cores of galaxies and exhibited light curves that look nothing like
SNe light curves: most rose or fell over periods of 100+ days. In
general, the continuous light curve monitoring in the survey made
possible to separate these artifacts and AGN from SNe with high
confidence.  For the remaining 29 candidates deemed to be SNe, we
determined a SN type and confidence for each. In Table~\ref{table:sne}
we list the SNe along with their host galaxy redshifts, SN types and
confidence. We omit the eight SNe whose hosts are spectroscopically
confirmed cluster members. See Figure~4 of B11 for images, light
curves and light curve fits of all candidates. A complete description
of the SN coordinates, typing and confidence level (plausible,
probable, or secure) is given in B11. Briefly, a secure SN~Ia has
either spectroscopic confirmation, or evidence from two sources
(early-type host galaxy and light curve) ruling out other types. A
probable SN~Ia is slightly less certain than a secure SN~Ia, but still
a high-confidence SN~Ia: the light curve rules out all core collapse
subtypes. A plausible SN~Ia has a light curve that is more indicative
of a SN~Ia than SN~CC, but is not sufficient to rule out all
core-collapse types.

In this analysis we use two additional selections not used in
B11: (1) First, we eliminate candidates that could only be consistent
with a SN~Ia if it peaked prior to 10 rest-frame days before the first
observation. We found that lower-redshift ($z \lesssim 0.9$) SNe were
detectable even when peaking well before the first observation, but
that such SNe were extremely difficult to type as they were observed
only far into the light curve decline. We found it most ``fair'' to
eliminate such candidates entirely. We include the same selection in
our efficiency simulations below. 
Only SCP06L21 and SCP05N10 are rejected based on this selection, but both are below the redshift range of greatest interest ($z>0.6$) and at least SCP05N10 is incompatible with an SN~Ia light curve anyway.
This was not an issue for B11 because SNe
of interest (at $z \ge 0.9$) are not detectable very far after peak.

(2) Second, we exclude regions within $20''$ of cluster centers, in
order to avoid the most strongly lensed areas in the volume behind the
clusters. This region is only $\sim$$3\%$ of the observed field of
each cluster. Note that we were careful to choose this radius before
looking at the radii of any of the candidates, in order to avoid
biasing ourselves by adjusting the radius to conveniently exclude or
include candidates. Two candidates were excluded as a result: SNe
SCP06B3 ($16.8''$ from the cluster center) and SCP06M50 ($19.4''$ from
the cluster center). As it happens, these candidates are unlikely to
be SNe~Ia. SN SCP06B3 is a ``probable'' SN~CC, while SN SCP06M50 is
possibly not a SN at all and may be hosted by a cluster member galaxy,
making its position near the cluster center unsurprising.  The
exclusion of this region is taken into account in our simulations
(\S\ref{ratecalc}). The effect of lensing on the remaining portions of
the fields are discussed in \S\ref{sec:lensing}.

\begin{figure*}[!tbh]
\epsscale{1.15}
\plottwo{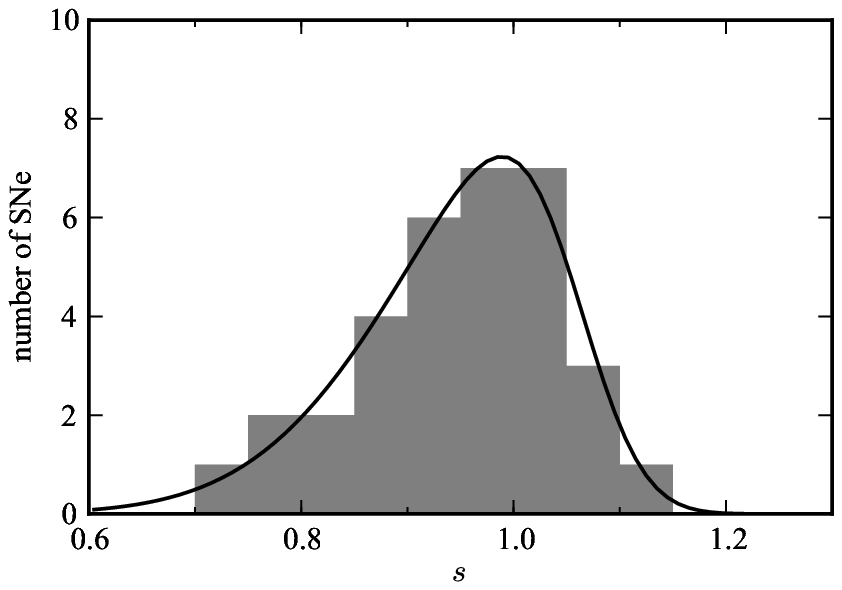}{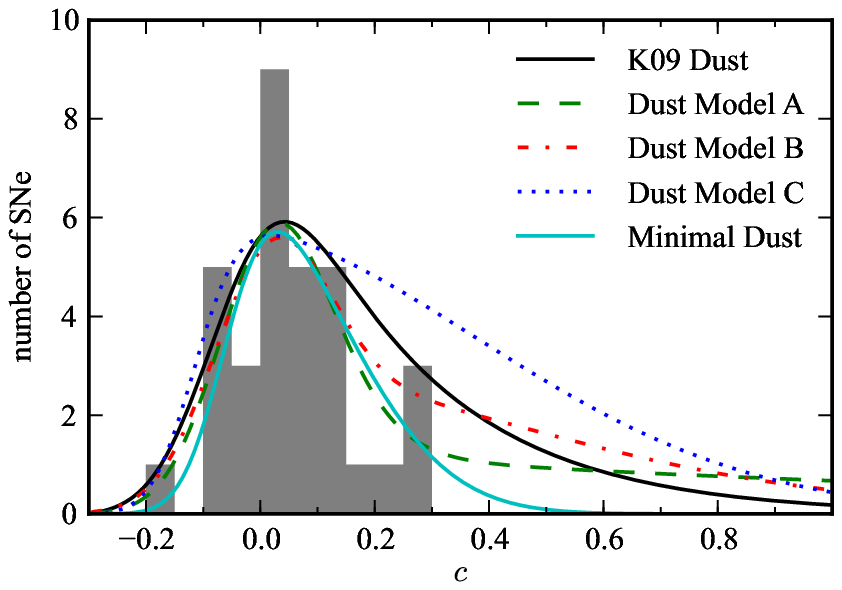}
\caption{\emph{Left Panel:} Stretch distribution used for simulated SNe
  (solid black line) and the stretch distribution of first-year
  SNLS $z<0.6$ SNe (grey histogram) from \citet{astier06a}.
  \emph{Right Panel:} Color distribution used for simulated SNe (solid
    black line), based on the K09 distribution of host-galaxy
  extinction.  The grey histogram shows the color distribution of the
  first-year SNLS $z<0.6$ SNe. The other four lines show alternative
  color distributions used to assess the possible systematic uncertainty due
  to different distributions of host galaxy dust extinction
  (see \S\ref{sec:sysdust}).
\label{fig:dists}}
\end{figure*}

The systematic uncertainties associated with the determination of SN
type and redshift for the remaining candidates are addressed
in \S\ref{sec:systyping}.

\vspace{0.1in}
\section{Rate Calculation} \label{ratecalc}

We calculate the SN~Ia rate in redshift bins using what has
become a standard method in rate calculations: The number of SNe~Ia
per unit time per comoving volume is estimated in the redshift bin
$z_1 < z < z_2$ by
\begin{equation} \label{eq:fieldrate}
\mathcal{R} (z_1 < z < z_2) = \frac{N_{\rm SN~Ia} (z_1 < z < z_2)}
        {\int_{z_1}^{z_2} T(z) \frac{1}{1+z}
          \frac{\Theta}{4\pi}\frac{dV}{dz}(z) dz}
\end{equation}
where $N_{\rm SN~Ia} (z_1 < z < z_2)$ is the number of SNe~Ia
discovered between redshifts $z_1$ and $z_2$, and the denominator is
the total effective time-volume for which the survey is sensitive to
SNe Ia in the redshift range $z_1<z<z_2$. $T(z)$ is the
\emph{effective visibility time} (also known as the ``control time'')
and is calculated by integrating the probability of detecting a SN~Ia
as a function of time over the active time of the survey. $T(z)$
depends on the dates and depths of observations, as well as the
specific requirements for selecting SNe. The factor of $1/(1+z)$
converts from observer-frame time to rest-frame time at redshift
$z$. The last two terms in the denominator represent the volume
comoving element between $z$ and $z + dz$ observed in the survey.
$\frac{dV}{dz}(z)$ is the comoving volume of a spherical shell of
width $dz$.  $\Theta$ is the solid angle observed in the survey, in
units of steradians. ($\Theta / 4\pi$ is the fraction of the spherical
shell we have observed.)  Finally, the average redshift of the bin,
weighted by the volume effectively observed, is given by
\begin{equation}
\bar{z} = \frac{\int_{z_1}^{z_2} z T(z) \frac{1}{1+z} \frac{dV}{dz}(z) dz}
        {\int_{z_1}^{z_2} T(z) \frac{1}{1+z} \frac{dV}{dz}(z) dz}.
\end{equation}

As in B11, we use an effective visibility time that depends on
position, as observation dates and depths vary within each observed
field. That is, in Equation~(\ref{eq:fieldrate}) we make the
substitution
\begin{equation}
T(z) \Theta \Rightarrow \int_{x,y} T (x,y,z) dx dy.
\end{equation}
$T(x,y,z)$ is calculated by simulating SN~Ia light curves at different
positions, redshifts and times during the survey, and determining the
probability that each simulated SN would be detected and counted in
our SN sample. We pass each simulated SN through the same automated
selections used to select the 60 candidates in our initial
sample. Additionally, we discard simulated SNe peaking prior to 10
rest-frame days before the first observation, as discussed in the
previous section.

We characterize the diversity of SN~Ia light curves as
a two-parameter family (stretch $s$ and color $c$) with an additional
intrinsic dispersion in luminosity. The absolute magnitude of each
simulated SN is set to
\begin{equation}
M_B = -19.31 - \alpha (s-1) + \beta c + I
\end{equation}
where $-19.31$ is the magnitude of an $s=1$, $c=0$ SN~Ia in our
assumed cosmology \citep{astier06a}, $\alpha = 1.24$, $\beta = 2.28$
\citep{kowalski08a}, and $I$ is an added ``intrinsic dispersion'',
randomly drawn from a Gaussian distribution centered at zero with
$\sigma = 0.15$~mag, as seen in \citet{kowalski08a}. This produces a
set of simulated SNe that closely matches the distribution seen in
\citet{kowalski08a}. To calculate the flux of each simulated SN in the
observed $z_{850}$ and $i_{775}$ filters, we use the \citet{hsiao07a}
spectral time series template.

The main difference from B11 is that we use distributions for stretch
and color that are representative of SNe in the field rather than in
clusters. For stretch, we look to the observed stretch distribution of
the first-year sample from the Supernova Legacy Survey
\citep[SNLS;][]{astier06a}, cut at $z<0.6$ to limit Malmquist bias
(Figure~\ref{fig:dists}, grey histogram). We assume this sample is
complete, fit a smooth curve to the distribution (same panel, solid
line), and use this for simulated SNe.  

For color, we cannot assume the SNLS sample (Figure~\ref{fig:dists},
histogram in right panel) is complete even at $z<0.6$, as highly
reddened SNe will have been missed. The standard picture today is that
the observed distribution of SN colors is due to a combination of both
intrinsic SN color variation and host galaxy extinction
\citep{guy10a,chotard11a}. Both of these are expected to introduce a
color that correlates with SN luminosity, possibly with different
strengths ($\beta$ is typically found to be smaller than the canonical
value of $R_B = 4.1$ for Milky Way dust). In order to capture both
effects with a single color distribution and a single $\beta$, we work
backwards from the desired host galaxy extinction distribution. We
wish to achieve a host galaxy extinction distribution of $P(A_V)
\propto \exp(-A_V/0.33)$ with $A_V \ge 0$, the best-fit value for
host-galaxy SN extinction in the SDSS-II SN Survey \citep[][hereafter
  K09]{kessler09a}.  To do this we use a color distribution of $P(c)
\propto \exp(-(\beta-1)c/0.33)$, because $A_V$ is related to $c$ via
$A_V = R_V \times E(B-V) \approx (\beta-1) \times c$. This ensures the
desired $A_V$ distribution is obtained for any given value of $\beta$.
The full $P(c)$ distribution is then a convolution of this $P(c)$
distribution from host galaxy dust and the intrinsic distribution of
SN color (assumed to be Gaussian). The Gaussian parameters of the
intrinsic distribution are chosen so that the full convolved $P(c)$
distribution matches the observed SNLS $c$ distribution at
$c<0.3$. The result is a $P(c)$ distribution (black line in right
panel of Figure~\ref{fig:dists}) that matches the SNLS sample where we
expect it to be complete ($c<0.3$) and also has the desired behavior
at large extinction based on our assumed $P(A_V)$ distribution. We use
this distribution in our simulation. In \S\ref{sec:sysdust} we assess
the systematic uncertainty associated with host galaxy dust by using
alternate color distributions obtained using the same method, but
different $P(A_V)$ distributions (other curves in the same panel).

$T(x,y,z)$ is calculated in bins of 100 $\times$ 100~pixels
($5''\ \times\ 5''$) in position. In each $5''\ \times\ 5''$ region,
we simulate 50 SN light curves with random parameters and random
position (within the bin) and take the average effective visibility
time of the 50 SNe ($\sim$80,000 SNe per field). Summing over all
areas observed in all 25 fields yields $T(z)$. In doing so we exclude
regions within $20''$ of cluster centers, as discussed in the previous
section. We calculate $T(z)$ at intervals of $\Delta z = 0.05$ in
redshift.

\section{Results and Systematic Uncertainty} \label{results}

\begin{figure}
\epsscale{1.175}
\plotone{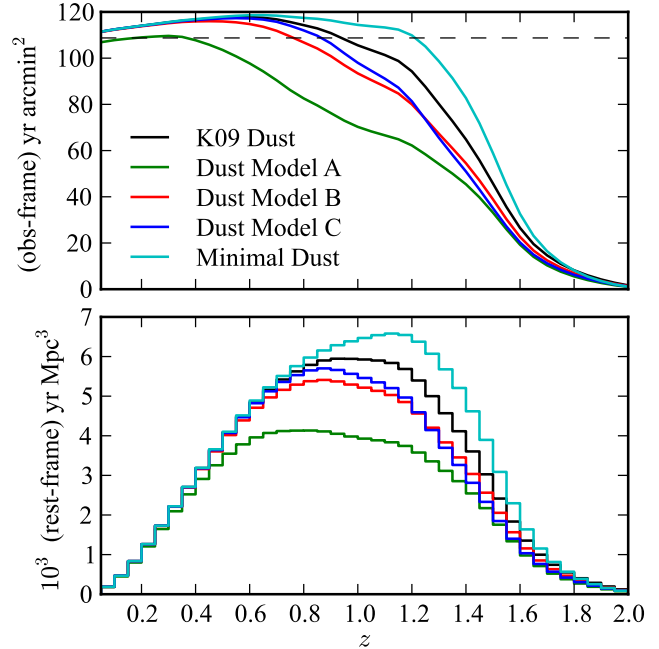}
\caption{\emph{Top Panel:} The observer-frame effective visibility time 
multiplied by observed area, as a function of supernova redshift. The
horizontal dotted line shows the area of the ACS field multiplied by
the time spanned by the observations in each cluster.
\emph{Bottom Panel:} The rest-frame volume-time searched in each redshift bin
of $\Delta z = 0.05$. In each panel, the black line shows our main
result for the effective visibility time, based on simulations using
the K09 dust distribution. The green, red, blue and cyan lines show
the results for alternative dust distributions. [\emph{Data in this
figure is available as a machine-readable table.}]\label{fig:ctarea1}}
\end{figure}

Figure~\ref{fig:ctarea1} (top panel, black line) shows the product of
the observer-frame effective visibility time and the area
($T(z) \Theta$ from Eq.~\ref{eq:fieldrate}) as a function of SN
redshift. For reference, the horizontal dotted line shows an
approximate calculation of this value, multiplying the area of the ACS
field (11.65~arcmin$^2$) by the time difference between 10~days before
the first observation and 10~days after the last observation. In
reality the area actually observed is slightly more complicated and
SNe are detected over a slightly larger time range. From $z=0$, the
effective visibility time actually increases slightly out to $z \sim
0.5$ as SN light curves are time-dilated and are thus visible for
longer. Afterwards, we begin to miss SNe that peak during the
observations. In the lower panel of Figure~\ref{fig:ctarea1}, we
convert to the rest-frame time-volume observed in each redshift bin of
$\Delta z = 0.05$ using the assumed cosmology.

Table~\ref{table:results} shows the results, in bins of $\Delta z =
0.4$ \citep[comparable to][]{dahlen08a} and also in bins of $\Delta z
= 0.5$ \citep[comparable to][]{graur11a}.  The numerator of
Equation~(\ref{eq:fieldrate}) (third column in
Table~\ref{table:results}) is the number of SNe~Ia described
in \S\ref{data}.  The denominator of Equation~(\ref{eq:fieldrate})
(fourth column in Table~\ref{table:results}) is obtained by summing
Figure~\ref{fig:ctarea1} (lower panel) over the redshift bin of
interest.  We now discuss the systematic uncertainties associated with
lensing, SN type determination, host-galaxy dust, SN properties, and
galaxy number density variations.

\subsection{Type determination} \label{sec:systyping}

The uncertainty in SN type in the survey is quite small, thanks to the
cadenced nature of the survey and excellent spectroscopic
follow-up. Consider the candidates designated as SN~Ia: all three
SNe~Ia at $z<0.9$ are spectroscopically confirmed. At $z \gtrsim 0.9$,
any SN bright enough to be detected is overwhelmingly likely to be
Type~Ia due to the faintness of core-collapse SNe relative to
SNe~Ia \citep[e.g.,][]{dahlen04a,li10a,meyers11a}. Furthermore,
while ``probable'' candidates are not as certain as ``secure''
candidates, this is still a fairly high-confidence type determination:
A ``probable'' SN~Ia means that a SN~Ia light curve template has a
$\chi^2$ $P$-value that is $10^3$ times larger than any SN~CC value. A
Bayesian analysis would therefore yield a type uncertainty close to
zero for such candidates, regardless of the prior used.

\begin{deluxetable}{lcccc}
\tablecaption{\label{table:results}Results: SN~Ia Rate}
\tablehead{\colhead{Redshift bin} & \colhead{$\bar{z}$} &
  \colhead{$N_{\rm SN~Ia}$\tablenotemark{a}} & 
  \colhead{Denom\tablenotemark{b}} & \colhead{Rate\tablenotemark{c}}}
\startdata
\cutinhead{bin width $\Delta z = 0.4$}
$0.2 < z \le 0.6$ & 0.442  & $0.00^{+1.16+0.00}_{-0.00-0.00}$ &2.332 & $0.00^{+0.50+0.00}_{-0.00-0.00}$ \\
$0.6 < z \le 1.0$ & 0.807  & $5.25^{+2.69+0.25}_{-2.00-1.25}$ &4.464 & $1.18^{+0.60+0.44}_{-0.45-0.28}$ \\
$1.0 < z \le 1.4$ & 1.187  & $5.63^{+2.77+0.63}_{-2.08-0.63}$ &4.243 & $1.33^{+0.65+0.69}_{-0.49-0.26}$ \\
$1.4 < z \le 1.8$ & 1.535  & $1.12^{+1.56+0.12}_{-0.79-1.12}$ &1.453 & $0.77^{+1.07+0.44}_{-0.54-0.77}$ \\
\cutinhead{bin width $\Delta z = 0.5$}
$0.0 < z \le 0.5$ & 0.357  & $0.00^{+1.16+0.00}_{-0.00-0.00}$ &1.624 & $0.00^{+0.71+0.00}_{-0.00-0.00}$ \\
$0.5 < z \le 1.0$ & 0.766  & $5.25^{+2.69+0.25}_{-2.00-1.25}$ &5.321 & $0.99^{+0.51+0.33}_{-0.38-0.24}$ \\
$1.0 < z \le 1.5$ & 1.222  & $6.75^{+2.99+0.75}_{-2.31-1.75}$ &4.906 & $1.38^{+0.61+0.71}_{-0.47-0.43}$ \\
$1.5 < z \le 2.0$ & 1.639  & $0.00^{+1.16+0.00}_{-0.00-0.00}$ &0.890 & $0.00^{+1.30+0.00}_{-0.00-0.00}$ 
\enddata

\tablenotetext{a}{Number of SNe~Ia in bin. The first and second confidence 
intervals represent the Poisson uncertainty and the uncertainty in
type determination, respectively. The non-integer number of SNe in
each bin is attributable to the two candidates without spectroscopic
redshifts. These candidates are assigned redshift ranges that are
spread over multiple bins.}
\tablenotetext{b}{Denominator of Equation (\ref{eq:fieldrate}): 
the total rest frame time-volume searched in this bin, having units
$10^4$ $h_{70}^{-3}$ yr Mpc$^3$.}
\tablenotetext{c}{The rate in units of $10^{-4}$ $h_{70}^3$ yr$^{-1}$ Mpc$^{-3}$. 
The first and second confidence intervals represent the statistical
and systematic uncertainty, respectively. The statistical uncertainty
is entirely due to Poisson uncertainty in $N_{\rm SN~Ia}$. The
systematic uncertainty is the typing uncertainty in $N_{\rm SN~Ia}$
and systematic uncertainties in ``Denom'' (described in text) added in
quadrature.}
\end{deluxetable}

\begin{deluxetable}{lcccccc}
\tablecaption{\label{table:syserr}SN Rate Uncertainties in Percentage}
\tablehead{ & & \emph{statistical} & &\multicolumn{3}{c}{\emph{systematic}}\\
\cline{3-3} \cline{5-7}
\colhead{\T Redshift bin} & & \colhead{Poisson} & & \colhead{Typing} & 
\colhead{Dust} & \colhead{Luminosity}}
\startdata
$0.6 < z \le 1.0$ & & +51/$-$38 & & +4/$-$23 & +37/$-$3 & +2/$-$2 \\
$1.0 < z \le 1.4$ & & +49/$-$36 & & +11/$-$11 & +49/$-$14 & +10/$-$7 \\
$1.4 < z \le 1.8$ & & +138/$-$69 & & +11/$-$100 & +39/$-$19 & +38/$-$23 
\enddata

\tablecomments{Percentages are not reported for the $0.2<z<0.6$ bin 
because there were no SNe detected in this bin.}
\end{deluxetable}

The ``plausible'' candidates are perhaps the only candidates with
significant type uncertainty. It is difficult to precisely quantify
this uncertainty. Instead, we provide conservative bounds in a manner
similar to \citet{dahlen08a} and \citet{sharon10a}: we first assign a
lower limit to the number of SNe~Ia discoveries by assuming that all
``plausible'' SNe~Ia are in fact SNe~CC.  We then assign an upper
limit by assuming that all ``plausible'' SNe~CC are in fact SNe~Ia.
These limits are shown in Table~\ref{table:results} as the second
confidence interval for $N_{\rm SN~Ia}$. The corresponding systematic
uncertainty in the SN rate is shown in Table~\ref{table:syserr}.

For the two candidates without spectroscopic host redshifts, we assign
a redshift range consistent with the SN light curve and/or host galaxy
photometry, as follows: For SCP06E12, we use the range $0.8<z<1.2$. As
there is uncertainty about both the type and cluster membership, we
count SCP06E12 as $0.5 \pm 0.5$ field SNe~Ia. The situation is similar
for SCP06N32: the light curve is consistent with an SN~Ibc at $z \sim
0.9$, but also with an SN~Ia at $z \sim 1.3$. We therefore assign a
redshift range of $1.1 < z < 1.5$ and count it as $0.5 \pm 0.5$ field
SNe~Ia.  These two SNe are assigned to redshift bins in
Table~\ref{table:results} according to the degree of overlap between
the redshift range of the SN and the range of the redshift bin.

Finally, note that we have classified the highest-redshift SN SCP06X26
as a lower-confidence ``plausible'' SN~Ia, despite the fact that any SN
detected at $z=1.44$ is overwhelmingly likely to be Type Ia. This
conservative approach was taken because the spectroscopic host galaxy
redshift of SCP06X26 is based on a single low signal-to-noise emission
line; as a result, the redshift and type are less certain than they
would otherwise be. The ``plausible'' designation therefore includes
the possibility that the SN is at a lower redshift. As a result, the
confidence interval on the number of detected SNe~Ia in the
$1.4<z<1.8$ bin is [0, 1]. Still, the light curve of SCP06X26 is
consistent with a typical SN~Ia at $z=1.44$, so this remains the best
estimate.

\subsection{Lensing due to Clusters} \label{sec:lensing}

The presence of a massive galaxy cluster in each of the 25 observed
fields presents a complication for measuring the volumetric field
rate. A cluster will preferentially magnify sources behind it
(increasing the discovery efficiency of SNe), and will also shrink the
source plane area $\Theta$ behind the cluster (decreasing the number
of SNe discovered) \citep[e.g.,][]{sullivan00a,goobar09a}. Fortunately
the effect of lensing on the calculated rates in this survey is small
for three reasons. First, the high redshifts of the clusters means
that the volume of interest in the cluster backgrounds is close to the
clusters and therefore not lensed very efficiently. Second, we have
already excluded from the analysis the central $20''$ of each field,
where lensing effects are the largest. Third, at any given radius the
two effects (magnification and source plane area shrinkage) are
opposing in terms of number of SNe discovered.

\begin{figure}
\epsscale{1.175}
\plotone{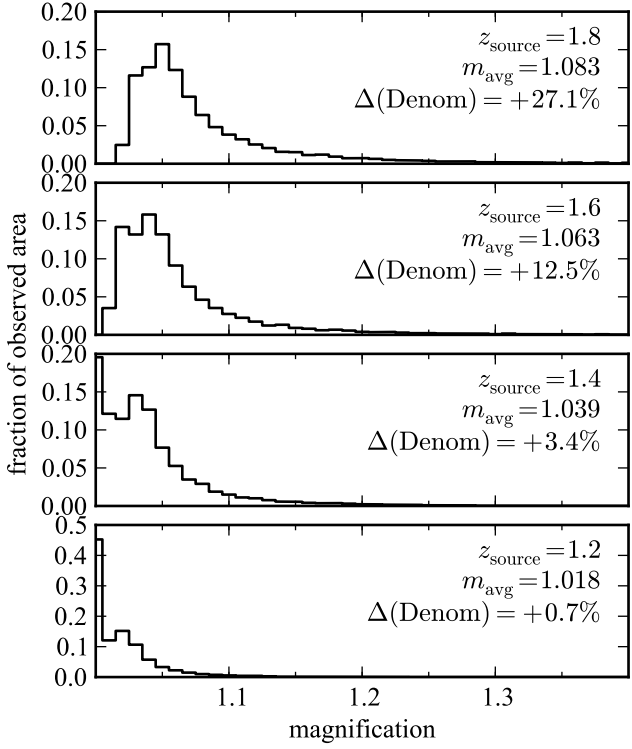}
\caption{Results of our lensing simulation: the magnification 
distribution in regions at radius $20'' < r < 100''$ in all 25 cluster
fields (the approximate extent of the regions used in the rate
analysis). The four panels correspond to source redshifts of $z_s =
1.2,\ 1.4,\ 1.6, 1.8$. For each source redshift, the average
magnification $m_{\rm avg}$ is given. Under the approximation that all
SNe are magnified by $m_{\rm avg}$, $\Delta({\rm Denom})$ shows the
corresponding increase in the denominator of Equation
(\ref{eq:fieldrate}). Note that this change in Denom considers only
magnification; the effect of source-plane area shrinkage is considered
separately in Figure~\ref{fig:sourcearea}. At $z<1.4$, where the survey's
sensitivity is greatest, magnification from lensing has only a
$\lesssim 3\%$ effect on the detectability of SNe.\label{fig:magpdf}}
\end{figure}

\begin{figure}
\epsscale{1.175}
\plotone{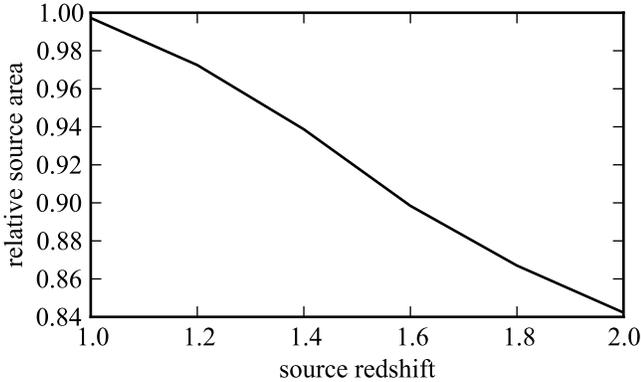}
\caption{True source-plane area relative to the observed area in our 
lensing simulation, as a function of source redshift. The relative
area is for regions at radius $20'' < r < 100''$ in each of the 25
cluster fields (the approximate extent of the regions used in the
analysis). The relative source area corresponds directly to the change
in the denominator of Equation~(\ref{eq:fieldrate}). The effect is
opposite in sign to the effect in Figure~\ref{fig:magpdf}.
\label{fig:sourcearea}}
\end{figure}

We have calculated the magnitude of each lensing effect on the
remaining outer regions using a simple lensing model: We assume each
cluster has a mass of $M_{200} = 4\times 10^{14}~M_\odot$ (the
approximate average mass in our sample, as reported by \citet{jee11a}
and an NFW \citep{navarro97a} mass profile. We distribute clusters
according to their redshifts and calculate the lensing effect on the
25 annular regions $20'' < r < 100''$ around the clusters. The
distribution of magnification in these regions as a function of source
redshift is shown in Figure~\ref{fig:magpdf}. The magnification is
quite small: even at a source redshift of $z=1.8$, most of the area is
magnified by less than 10\%. As a rough estimate of the effect on the
derived rates, we show the average magnification $m_{\rm avg}$ for
each source redshift, and the effect such a magnification would have
on the detectability of SNe at this redshift. To calculate the effect
on the detectability, we increase the luminosity of all SNe in our
Monte Carlo efficiency simulation by a factor $m_{\rm avg}$ and
recalculate the denominator of Equation~(\ref{eq:fieldrate}). More
luminous SNe causes ``Denom'' to increase, corresponding to a decrease
in the inferred SN rate. The effect is only a few percent at
$z \lesssim 1.4$, where the survey is most sensitive, but starts to
increase steeply towards $z=1.8$. In Figure~\ref{fig:sourcearea} we
show the decrease in the source area as a function of redshift, which
translates directly into a decrease in the effective visibility time
$\times$ area. The decrease is nearly linear with redshift past
$z=1.2$, reaching $\sim$13\% at $z=1.8$.

We conclude from these simulations that the two effects cancel to
within a few percent of the total rate, over the redshift range of
interest: At $z=1.4$, magnification increases SN detectability by
$\sim$3\% and source-area reduction decreases detectability by
$\sim$6\%. At $z=1.6$, the increase is $\sim$12\%, and the decrease is
$\sim$10\%. At $z=1.8$ the increase overwhelms the decrease
($\sim$27\% versus $\sim$13\%), but there will be very few SNe
detected beyond $z \sim 1.6$ (see Figure~\ref{fig:ctarea1}). Therefore,
we have not made an adjustment for these effects. Furthermore, the
size of each effect is much smaller than other sources of systematic
uncertainty considered below. For example, the average magnification at
$z=1.8$ is only $\sim$1.08 ($-0.08$~mag), whereas below we consider
the effect of changing the luminosity of all SNe in our simulation by
$\pm 0.2$~magnitudes. So, we do not assign a specific systematic uncertainty
to lensing effects.

\begin{figure}
\epsscale{1.175}
\plotone{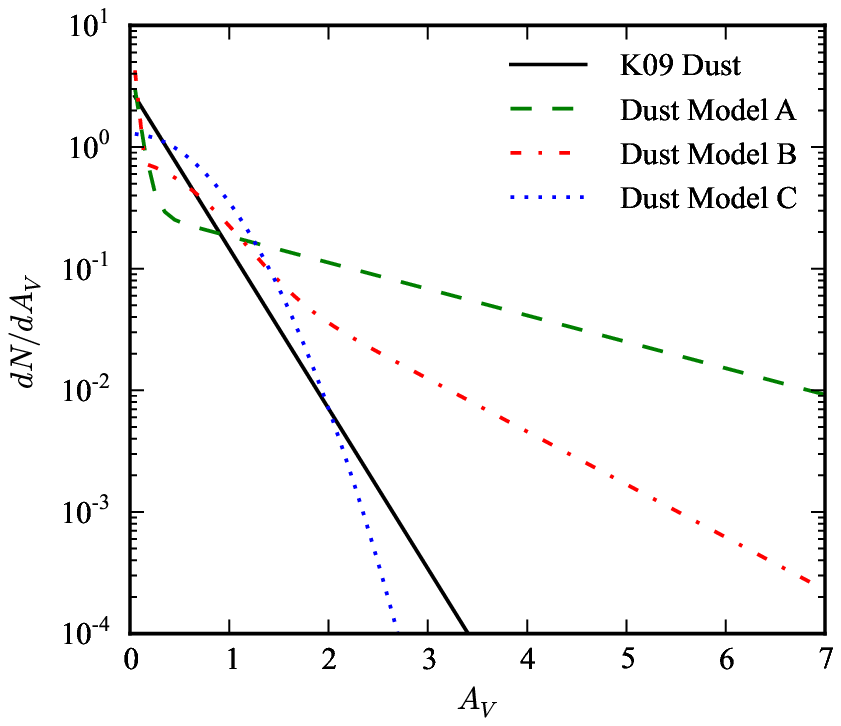}
\caption{Host galaxy dust extinction distributions, illustrating the 
  behavior of the distributions at large $A_V$. The K09 distribution is
  used for our main result. Models A, B, and C are similar to the
  models of the same name examined in \citet{dahlen08a} and are based
  on results from \citet{hatano98a}, \citet{riello05a}
  and \citet{neill06a}, respectively. These alternative distributions
  are used here to investigate possible systematic uncertainty due to host
  galaxy dust. This figure can be compared to Figure~3
  of \citet{dahlen08a}.\label{fig:extinction}}
\end{figure}

\begin{figure*}
\epsscale{1.15}
\plottwo{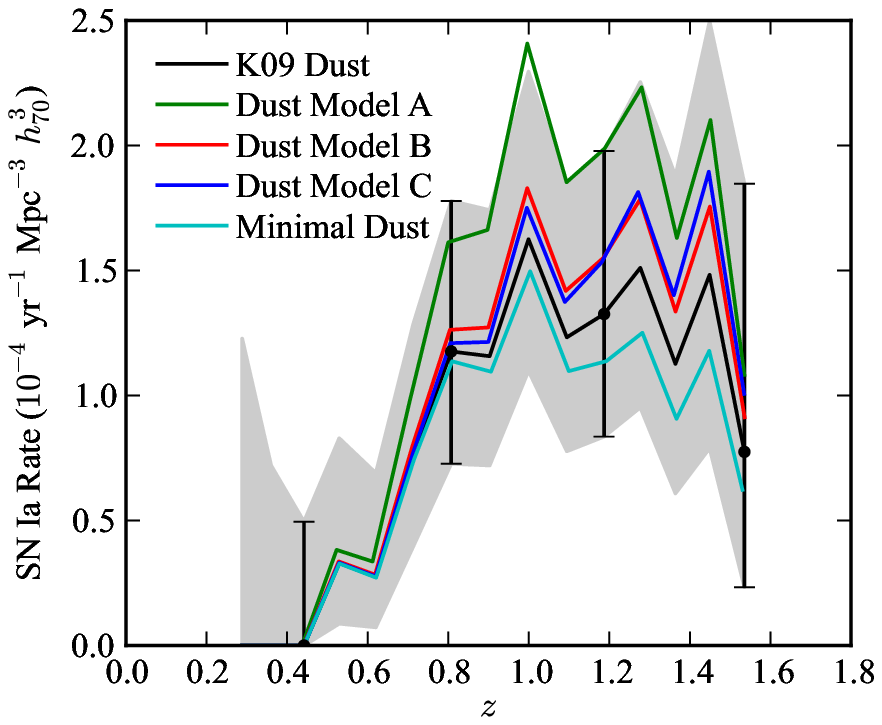}{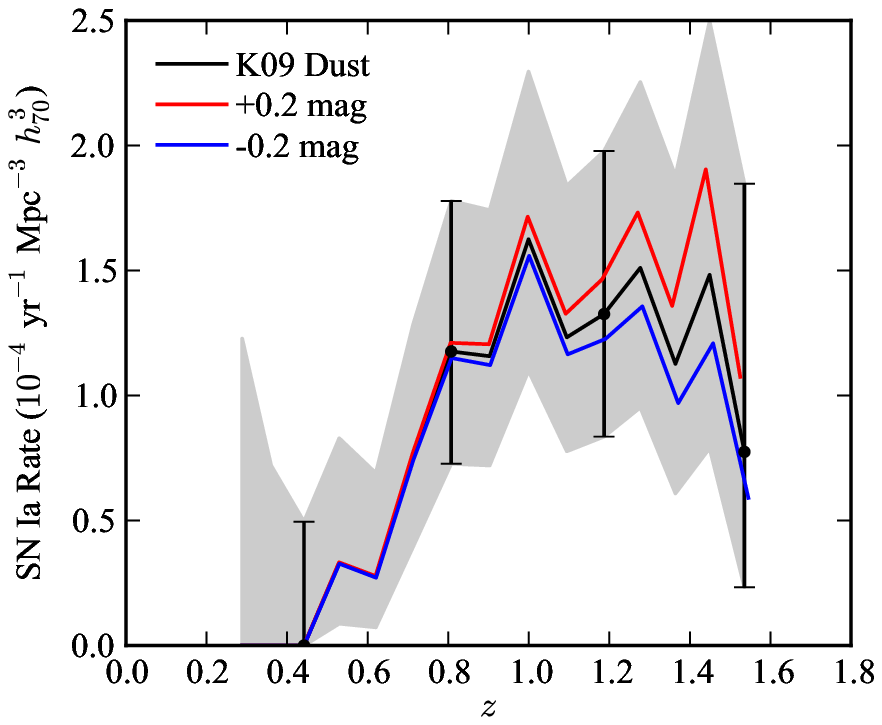}
\caption{The volumetric SN Ia rate in four redshift bins 
(points with error bars) of width $\Delta z = 0.4$. The error bars
  represent the statistical-only uncertainty. The black line shows the rate
  calculated in a moving bin of width $\Delta z = 0.4$ (shaded grey regions
  represent uncertainty). Note that the points with error bars are
  uncorrelated errors (using non-overlapping bins), while the
  uncertainty in the moving bin is correlated from point to point.
\emph{Left Panel}: The green, red, blue and cyan lines show the 
rate (with no uncertainty) assuming alternative SN color distributions.
\emph{Right Panel}: The red and blue lines show the rate assuming 
that all SNe are brighter or dimmer by $\pm 0.2$~mag.
\label{fig:fieldrates1}}
\end{figure*}

\subsection{Dust Extinction} \label{sec:sysdust}

The degree to which SNe are affected by host galaxy dust extinction is
perhaps the largest systematic uncertainty in SN~Ia rate
studies. Here, we consider alternatives to the extinction distribution
used in \S\ref{ratecalc} and evaluate the effect on the
results. Specifically, we reproduce three $P(A_V)$ distributions
considered in \citet{dahlen08a} and \citet{neill06a}.  In each of
these models, as for our main result, we constrain $P(A_V) = 0$ for
$A_V < 0$.

The first, ``Model A,'' is used for the main
result in \citet{dahlen08a}. It is based on the model
of \citet{hatano98a}, constructed to estimate extinction in local disk
galaxies. The distribution in Figure 3 of \citet{dahlen08a} is
well approximated by
\begin{equation}
P(A_V) = 0.61\frac{e^{-A_V/2}}{2} + 0.39\frac{e^{-A_V/0.07}}{0.07}.
\end{equation}
This distribution is shown in Figure~\ref{fig:extinction}. The second
distribution we consider, ``Model B,'' is used by \citet{dahlen08a} as
an alternative distribution. It is based on the models
in \citet{riello05a}, which are aimed at generalizing
the \citet{hatano98a} models to a variety of dust properties and a
variety of spiral Hubble types. Here we approximate the distribution
by
\begin{eqnarray}
P(A_V) &=& 0.35 \delta(A_V) +0.40\frac{2}{0.6 \sqrt{2\pi}}e^{-A_V^2/(2 \times 0.6^2)} + \nonumber \\
       & & +~0.25 e^{-A_V}.
\end{eqnarray}
where $\delta$ is the Dirac delta function, used here to assign 35\%
of SNe to the lowest host galaxy extinction bin.  The third
distribution we consider, labeled ``Model C,'' is used in the rate
analysis of \citet{neill06a}. It is given by
\begin{equation}
P(A_V) = \frac{2}{0.62 \sqrt{2\pi}}e^{-A_V^2/(2 \times 0.62^2)}.
\end{equation}
All three of these distributions are reproduced in
Figure~\ref{fig:extinction}, and the corresponding distributions of SN
color are shown in Figure~\ref{fig:dists} (right panel).  In addition
to these three distributions, we also consider a distribution with
minimal dust, where we assume the SNLS $z<0.6$ sample is complete and
fit it with a skewed Gaussian distribution.  The effective visibility
time for each dust model is shown in Figure~\ref{fig:ctarea1} and the
corresponding SN rate results are shown in
Figure~\ref{fig:fieldrates1} (left panel).

Of all the models, Model A produces the most strikingly different
results for the effective visibility time. Even in the lowest redshift
bin ($0.2<z<0.6$) it implies that 10\% of SNe are missed due to dust,
relative to the K09 model. In the $0.6<z<1.0$ redshift bin, it yields
effective visibility times lower by 27\%, while Models B, C and the
minimal dust model result in changes of only $-7\%$, $-3\%$, and
$+3\%$ respectively. This is unsurprising: In Model A, 26\% of
SNe have host galaxy extinctions $A_V > 2$ while this fraction is
$<4\%$ in Model B and $<1\%$ in models K09 and C. In the $1.0 < z <
1.4$ bin Model A has the largest effect: $-33\%$ compared to the K09
model. As the work of \citet[][Model B]{riello05a} is aimed at
generalizing \citet[][Model A]{hatano98a} for use at higher redshifts,
Model B may be viewed as more applicable. Along these
lines, \citet{cappellaro99a} have noted that using
the \citet{hatano98a} model appears to over-correct SN rates.

Still, for the systematic uncertainty associated with the choice of
dust model we take a conservative approach, using the full range of
these models. We use the minimal dust model to obtain the lower limit
on the rate and Model A for the upper limit. This confidence range is
shown in Table~\ref{table:syserr}.

\subsection{More Dust at High Redshift?} \label{sec:hizdust}

Several studies have pointed out that extinction is likely to increase
with redshift, due to increasing star formation in dusty
environments \citep{mannucci07a,holwerda08a}. The potential effect on
the SN~Ia extinction distribution is difficult to quantify as it
depends not only on the relation between star formation and dust
ejection but also the SN~Ia delay time distribution. However, we
estimate that any possible redshift-dependence is well-encompassed by the
conservatively large range of extinction models we use above.
\citet{rowanrobinson03a} estimated that the average extinction $\langle
A_V \rangle (z)$ peaks at $z \sim 1$, at a value $\sim$0.15~mag higher
than locally. The effect of such an additional dimming on our rates at
$z \sim 1$ is approximately 10\%, whereas the extinction distribution
uncertainty already included above is
$\sim$50\%. Similarly, \citet{graur11a} estimated the fraction of
missing SNe to be 5--13\% in the redshift range $1<z<2$ based on the
work of \citet{mannucci07a}. As the difference induced by our use of
the K09 extinction model rather than model C used by Graur et al. is
already much greater than this, we do not make an explicit correction
for increasing dust at high redshift.

\subsection{Other SN properties}

In addition to the choice of the host galaxy dust extinction
distribution, other assumptions about SN properties can affect the
derived rates.  These include the absolute magnitude, the stretch
distribution, and the spectral time series template. Fortunately these
properties are well constrained. For example, shifting the stretch
distribution by $\Delta s = \pm 0.05$ would be inconsistent with the
observed distribution, and similarly for the color distribution. The
average absolute magnitude of SNe~Ia (in our cosmology) is constrained
to much better than $0.1$~mag \citep{amanullah10a}.  To first order,
changing any of these assumptions affects the derived rate by
increasing or decreasing the luminosity of the simulated SN and
thereby increasing or decreasing the effective visibility
time. Therefore, to estimate the effect of such changes on the result,
we simply shift the absolute magnitude of the simulated SNe~Ia. A
shift in stretch of $\Delta s = \pm 0.05$ is equivalent to a magnitude
shift of $\Delta M_B = \alpha\Delta s = 0.08$. Similarly, a shift in
color of $\Delta c = \pm 0.05$ is equivalent to a magnitude shift of
$\Delta M_B = \beta\Delta c = 0.11$. Conservatively then, we use a
$\pm 0.2$~mag shift to jointly capture these uncertainties and
uncertainty in the absolute magnitude. The effect on the results is
shown in Figure~\ref{fig:fieldrates1} (right panel) and is generally
comparable to or smaller than the effect from the extinction
distribution uncertainty (see Table~\ref{table:syserr}).

\subsection{Cosmic Variance and galaxy-cluster correlation} \label{sec:cv}

There are various effects that could increase or decrease the density
of galaxies in the observed fields relative to the cosmic average,
potentially biasing the volumetric rates. We have estimated all such
effects to be small ($\lesssim 5\%$). We briefly discuss three of
these effects.

(1) \emph{Masking volume surrounding clusters.}  Any SN occurring in
the volume around a cluster (within $|\Delta z| \lesssim 0.015$) but
not associated with the cluster would be mistakenly assigned to the
cluster.  This volume is therefore effectively not counted for this
analysis. We estimate that this decreases the total volume in the
$0.6<z<1.0$ redshift bin by $\approx$1.5\% (due to the presence of 5
clusters) and in the $1.0<z<1.4$ bin by $\approx$6\% (due to the
presence of 20 clusters). Because the next effect compensates somewhat
for this effect, we do not make an explicit correction in the result.

\begin{figure*}
\epsscale{1.}
\plotone{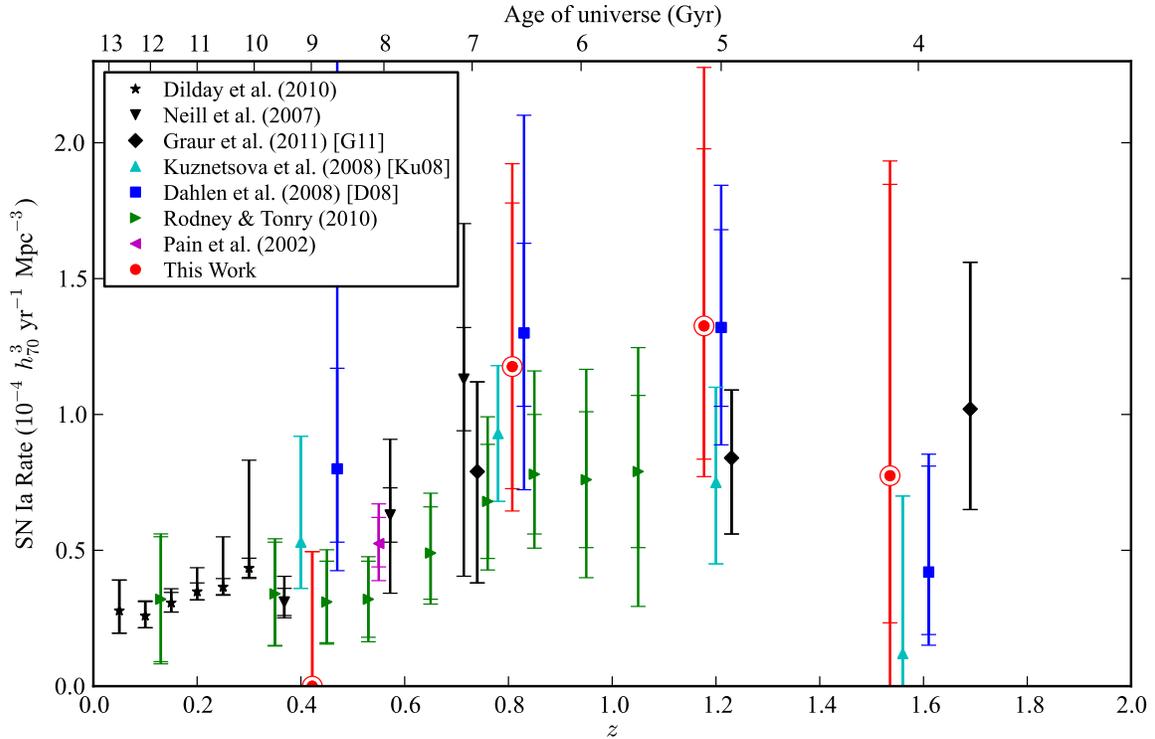}
\caption{Volumetric SN~Ia rates from the \emph{HST} Cluster Supernova
  Survey (red points) compared to key rates from the literature. For
  measurements with two error bars, including ours, the inner and
  outer error bars represent the statistical (Poisson) and total
  (statistical + systematic) uncertainties, respectively. Measurements
  with a single error bar (Ku08 and G11) are Bayesian-based analyses
  where the error bar encompasses both statistical and typing
  uncertainties. Some measurements have been offset slightly in
  redshift for visibility. Note that we have conservatively reported
  our measurement in the highest-redshift bin as having a lower limit
  of zero due to redshift uncertainty in the single SN in this
  bin.\label{fig:ratecompilation}}
\end{figure*}

(2) \emph{Cluster-galaxy correlation.}  There may be an increase in
the density of field galaxies (relative to the cosmic average) in the
volume surrounding the cluster due to the presence of the cluster
itself. However, we estimate that this effect should be small outside
of the volume discussed in the previous effect. At $z=1$, a redshift
difference of $\Delta z=0.015$ corresponds to a comoving distance of
$\sim$36~Mpc. Density fluctuations on these scales are in the linear
regime. At 36~Mpc, the correlation function of massive galaxies is
$\xi \lesssim 0.2$ \citep[e.g.,][]{eisenstein05a,white11a}. Even if
the average over-density at scales 36~Mpc $<r<$ 100~Mpc were 0.2 (an
overestimate), this would only be a 2--3\% effect on the total galaxy
density in a bin of width $\Delta z = 0.4$.

(3) \emph{Cosmic variance.}
Cosmic variance could affect the results if there is an under-density
or over-density of field galaxies in the observed regions (relative to
the cosmic average). In B11, we estimated the size of this effect in
the GOODS fields, finding it to contribute a systematic uncertainty of
4--6\% based on the cosmic variance calculator of \citet{trenti08a}
and the difference between the GOODS-N and GOODS-S fields \citep[see
also discussion in][]{dahlen08a}. The effect is even smaller for the
25 fields in this study: here we have more than two widely separated fields,
providing a better sampling of cosmic variance.

\section{Discussion} \label{discussion}

The results from this study are available as a machine-readable table
from the Journal or the survey
website\footnote{\url{http://supernova.lbl.gov/2009ClusterSurvey/}}. We
include the product of the effective visibility time and observed
area, as a function of redshift (Figure~\ref{fig:ctarea1}), for a
variety of assumptions about SN properties and host galaxy dust
distributions. With these data and the SN candidate list, the rates
from this survey can be recomputed in any arbitrary redshift bin and
for any of these assumptions. This will make it easy to combine these
results with other measurements for increased statistical power.

\subsection{Comparison to Other High-Redshift Measurements}

In Figure~\ref{fig:ratecompilation} we compare our results to an
assortment of other volumetric SN Ia rate measurements. (All
measurements have been corrected to our assumed cosmology.) Our
results are generally consistent with the three published measurements
at $z \gtrsim 1$: \citet[][hereafter
Ku08]{kuznetsova08a}, \citet[][hereafter D08]{dahlen08a}, and
\citet[][hereafter G11]{graur11a}. D08 and G11 supplant earlier
results from \citet{dahlen04a} and \citet{poznanski07a},
respectively. The Ku08 and D08 measurements are based on SN searches
in the \emph{HST} GOODS fields, with Ku08 being an independent
analysis of a subset of the data used in D08. These SN searches used
ACS to cover the GOODS fields with a 45 day cadence and triggered
followup (imaging and spectroscopy) of SN candidates. The D08 analysis
uses a SN typing method based on both spectroscopy and photometry
(similar to the approach used here) while Ku08 use a photometric-only
pseudo-Bayesian approach to typing. The G11 measurement is based on
``single-detection'' searches in the Subaru Deep Field. G11 also use a
pseudo-Bayesian typing approach, but use a single detection with
observations in three filters, rather than multiple detections with
observation in (typically) two filters as in Ku08.

\begin{deluxetable*}{p{1.6in}ccc}
\tablewidth{4.55in}
\tablecaption{\label{table:comparerates}Rate Comparison Using 
Consistent Extinction Distributions}
\tablehead{\colhead{} & \colhead{$0.6 < z < 1.0$} &
  \colhead{$1.0 < z < 1.4$} & \colhead{$1.4 < z < 1.8$}}
\startdata
This work using extinction model A  & $1.61^{+0.83}_{-0.62}$  & $1.99^{+0.98}_{-0.74}$  & $1.08^{+1.50}_{-0.76}$ \\
\B D08 & $1.30^{+0.33}_{-0.27}$ & $1.32^{+0.36}_{-0.29}$ & $0.42^{+0.39}_{-0.23}$ \\
\hline
\T This work using minimal extinction model  & $1.14^{+0.58}_{-0.43}$  & $1.14^{+0.56}_{-0.42}$  & $0.62^{+0.86}_{-0.44}$ \\
\B Ku08 & $0.93^{+0.25}_{-0.25}$ & $0.75^{+0.35}_{-0.30}$ & $0.12^{+0.58}_{-0.12}$ \\
\hline
\hline
\T \B & $0.5<z<1.0$ & $1.0<z<1.5$ & $1.5<z<2.0$ \\
\hline
\T This work using extinction model C and +5 -- 13\% high-$z$ correction & $1.01^{+0.52}_{-0.39}$  & $1.69^{+0.75}_{-0.58}$  & $0.00^{+1.92}_{-0.00}$ \\
G11 & $0.79^{+0.33}_{-0.41}$ & $0.84^{+0.25}_{-0.28}$ & $1.02^{+0.54}_{-0.37}$
\enddata
\tablecomments{Rate in units of $10^{-4}$ $h_{70}^3$ yr$^{-1}$ Mpc$^{-3}$. 
Confidence intervals are statistical (Poisson).}
\end{deluxetable*}

\subsection{The Effect of Different Extinction Distributions}

Although our results generally agree with all three previous studies
(Ku08, D08, G11), it is interesting to compare in more detail the
assumptions used in each study.  Here we focus specifically on
differences in the assumed extinction distribution, which we find to
be the dominant systematic.  Unlike systematic uncertainties arising
from SN typing, systematic uncertainty due to assumptions about SN
properties will be common between rate studies, potentially leading to
a systematic offset between results if different assumptions are used
in different studies.

For example, our results appear very similar to D08 in the two
mid-redshift bins (within 10\% in both bins). However, D08 assume a
different extinction distribution in their simulations than we do. If
we assume the same distribution (extinction model A), our derived
rates in these two bins are 1.61 and 1.99 $\times
10^{-4}$~$h_{70}^{3}$, compared to 1.30 and 1.32 $\times
10^{-4}$~$h_{70}^{3}$ in D08. (Conversely, if D08 had used the K09
extinction model, their rates would have been lower than ours.)  In
other words, we actually found more SNe~Ia than one would have
predicted based on D08 (but not inconsistently so given the large
Poisson uncertainties).

Note that D08 also compare Models A, B and C in assessing their
systematic uncertainty but do not find the large differences that we
find here. They find that Model B produces rates that are $\lesssim
10\%$ lower than Model A (only $\sim$$4\%$ in the highest redshift
bin), and that Model C has even less of an effect, whereas we find
that the difference beyond Models A and B is approximately 26\% in the
central two bins. These different findings regarding the systematic
uncertainty due to extinction are surprising given that both studies
are based on HST searches the same band to similar depths. One
possible explanation is the difference in cadence ($\sim$23~days here
versus $\sim$45~days in GOODS). We checked the impact of a longer
cadence by rerunning our simulations ignoring every other epoch in our
search.  We find that the effect of different extinction distributions
is unchanged to within a few percent. That is, the shorter
$\sim$23~day cadence is not significantly more sensitive to highly
reddened SNe than the $\sim$45~day cadence. The slight enhancement in
reach that a short cadence affords is insignificant compared to the
long $A_V$ tail of, for example, Model A. It is possible that the
difference in estimated systematic uncertainty could arise from some
other detail of the efficiency simulations, but a full resimulation of
the D08 efficiency calculations is beyond the scope of this paper.

In Table~\ref{table:comparerates} we have recomputed our rates using
extinction assumptions similar to D08, Ku08 and G11.  Ku08 use only
two discrete values for $A_V$. The values, 0.0 and 0.4, are relatively
small, so we compare using our ``minimal dust'' model. G11 use the
distribution of N06 (Model C) with a $+$5--13\% correction in the
redshift range $1.0 < z < 2.0$ as discussed earlier (\S\ref{sec:hizdust}).
Table~\ref{table:comparerates} serves two purposes: (1) it aids
direct comparison between each of these studies and our
result, and (2) it illustrates how much the assumed extinction
distribution affects our result.

In light of the large systematic differences due to dust, it is
vitally important to use caution when comparing rate results from
studies that use different dust assumptions. In particular, systematic
offsets from dust assumptions could affect the shape of the derived
SN~Ia delay time distribution. The DTD shape is obtained not from the
SN rate itself, but from the change in the rate with
redshift \citep[see, for example, Figure 2 of][]{horiuchi10a}. To
measure the change, one typically must compare surveys covering
different redshift ranges. Comparing low- and high-redshift
measurements that correct their rates using different extinction
distributions will induce a systematic error on the slope of the SN
rate and thereby the DTD shape. 

To avoid such a systematic bias, studies of the DTD should strive to
use consistent extinction corrections between low and high
redshift. To aid this, we have provided our rates calculated under a
variety of extinction assumptions. Consitency will go a long way
towards reducing potential errors in the DTD, even if the extinction
distribution remains poorly known. However, in the long run one would
like a better understanding of SN extinction distributions at
both low and high redshift, particularly to avoid uncertainties due to
a \emph{changing} extinction distribution with redshift.  The
prospects for directly detecting the highly extincted SNe are not
great: even with a deep IR search, most SNe with $A_V>2$ will be
missed at high redshift. The alternative is more precise updated modeling of
SN~Ia extinction in the vein of \citet{riello05a}
or \citet{mannucci07a} that takes into account factors such as the
evolution of extinction with delay time and our latest knowledge of
the SN~Ia DTD.

\subsection{Summary}

In this paper we have computed volumetric SN~Ia rates based on 189
\emph{HST} orbits. This large \emph{HST} dataset adds statistics to
the existing \emph{HST} rate measurements, previously based only on
the GOODS fields. Our results provide additional strong evidence that
the SN~Ia rate is $\gtrsim 0.6 \times 10^{-4}$~$h_{70}^{3}$ yr$^{-1}$
Mpc$^{-3}$ at $z \sim 1$. The availability of raw data from our
efficiency simulations makes it simple to combine this dataset with
current and future \emph{HST} datasets, such as the 902-orbit Cosmic
Assembly Near-IR Deep Extragalactic Legacy Survey
\citep[CANDELS;][]{grogin11a}, for even greater statistical gains.

We find that an important systematic uncertainty in our result is the
amount of host-galaxy dust assumed in our simulations. 
This illustrates the need to use caution in comparing SN rate results
from different surveys, especially as statistical uncertainty
decreases and systematic uncertainties become dominant. Consistent
comparisons and updated extinction models can drastically reduce the
dust systematic in future studies.

\acknowledgements
Financial support for this work was provided by NASA through program
GO-10496 from the Space Telescope Science Institute, which is operated
by AURA, Inc., under NASA contract NAS 5-26555.  This work was also
supported in part by the Director, Office of Science, Office of High
Energy and Nuclear Physics, of the U.S. Department of Energy under
Contract No. AC02-05CH11231, as well as a JSPS core-to-core program
``International Research Network for Dark Energy'' and by a JSPS
research grant (20040003).  The authors wish to recognize and
acknowledge the very significant cultural role and reverence that the
summit of Mauna Kea has always had within the indigenous Hawaiian
community.  We are most fortunate to have the opportunity to conduct
observations from this mountain.  Finally, this work would not have
been possible without the dedicated efforts of the daytime and
nighttime support staff at the Cerro Paranal Observatory.
{\it Facilities:} \facility{HST (ACS)}, \facility{Subaru
     (FOCAS)}, \facility{Keck:I (LRIS)}, \facility{Keck:II
     (DEIMOS)}, \facility{VLT:Antu (FORS2)}

\end{document}